\documentstyle[12pt]{article}
\newcommand{\NP}[3]{Nucl.\ Phys.\ {\bf B#1} (#2) #3}
\newcommand{\PL}[3]{Phys.\ Lett.\ {\bf B#1} (#2) #3}
\newcommand{\PR}[3]{Phys.\ Rev.\ {\bf D#1} (#2) #3}
\newcommand{\PRL}[3]{Phys.\ Rev.\ Lett.\ {\bf #1} (#2) #3}
\newcommand{\ZP}[3]{Z.\ Phys.\ {\bf C#1} (#2) #3}
 
\newcommand{\IP}{{I\hspace{
-1mm}P}}
\newcommand{\IPs}{{I\hspace{-0.5mm}P}}

\newcommand{\simgt}{\,\rlap{\lower 3.5 pt \hbox{$\mathchar \sim$}} \raise 1pt
 \hbox {$>$}\,}
\newcommand{\simlt}{\,\rlap{\lower 3.5 pt \hbox{$\mathchar \sim$}} \raise 1pt
 \hbox {$<$}\,}
\def \bea   {\begin{eqnarray}}
\def \eea   {\end{eqnarray}}
\def \beq   {\begin{equation}}
\def \eeq   {\end{equation}}

\newcommand{\gev}{\,\mbox{GeV}}

\begin{document}
\title{\vskip-3cm{\baselineskip16pt
\centerline{\normalsize DESY 94--140\hfill ISSN 0418-9833}
\centerline{\normalsize KEK--TH--407\hfill}
\centerline{\normalsize KEK Preprint 94--77\hfill}
\centerline{\normalsize hep--ph/9408340\hfill}
\centerline{\normalsize August 1994\hfill}}
\vskip1.5cm
Diffractive Photoproduction of Jets with a Direct Pomeron Coupling at HERA}
\author{B.A. Kniehl\thanks{Present Address: KEK Theory Group, 1--1 Oho,
Ibaraki--ken, Tsukuba 305, Japan; address after 1~October 1994:
Max--Planck--Institut f\"ur Physik, F\"ohringer Ring 6, 80805 Munich,
Germany.},\mbox{\,}\thanks{Supported by Japan Society for the Promotion
of Science (JSPS) through Fellowship No.~S94159.}\mbox{\ }
H.--G. Kohrs\thanks{Supported by Deutsche Forschungsgemeinschaft (DFG)
through Graduiertenkolleg.}
\mbox{\ } and G. Kramer\\
II. Institut f\"ur Theoretische Physik\thanks{Supported
by Bundesministerium f\"ur Forschung und Technologie (BMFT), Bonn, Germany,
under Contract 05~6~HH 93P~(5)
and by EEC Program {\it Human Capital and Mobility} through Network
{\it Physics at High Energy Colliders} under Contract
CHRX--CT93--0357 (DG12~COMA).},\mbox{\ } Universit\"at Hamburg\\
Luruper Chaussee 149, 22761 Hamburg, Germany}
\date{}
\maketitle
\begin{abstract}
 
We investigate in detail the effect of a direct pomeron coupling
to quarks on the production of jets in $ep$ scattering with almost real
photons.
Jet production via a direct pomeron coupling is compared with
the resolved--pomeron mechanism.
We consider both direct and resolved photoproduction.
Rapidity and transverse momentum distributions
are calculated and compared with preliminary H1 and ZEUS data.
 
\end{abstract}
\newpage
 
\parindent6mm
\section{Introduction}
\mbox{}
 
The production of high--transverse--momentum jets by
quasi--real photons on protons
is one of the major processes to gain further insight into the interactions
of photons with quarks and gluons. Experimental results by the H1 \cite{H1A}
and ZEUS \cite{ZEUSA} Collaborations at HERA have been presented, and more
and better data are expected to come out soon. A subsample of these events
are so--called {\it large--rapidity--gap events}, which have been discovered
recently in photoproduction and deep--inelastic electroproduction in both
HERA experiments \cite{ZEUSB,H1B}. Their properties were found to be
inconsistent with the dominant dynamical mechanism for the production of jets
in photoproduction and deep--inelastic electroproduction, where colour is
transferred
between the produced quark and gluon jets and the proton remnant. Whereas
the majority of the events clustered around $\eta_{max}=4$, a second
class of events, with $\eta_{max} \le 1.5$, were observed, i.e., they
had a large gap in rapidity between the fast moving proton,
which in the moment is still an assumption,
and the rest of the hadronic final state.
(Here, $\eta$ is the rapidity measured along the
incoming--proton direction in the laboratory frame.)
The same type of events had been observed
already some time ago in $p\overline{p}$ scattering \cite{UA8A}.
They were interpreted as being due to diffractive hard scattering, following a
suggestion of Ingelman and Schlein \cite{IS}. In this interpretation,
which also applies to the large--rapidity--gap events at HERA, the proton
emits a pomeron which acts as a virtual target for the incoming hadron
(in the case of $p\overline{p}$ scattering) and the quasi--real or highly
virtual photon (in the case of $ep$ scattering), respectively.
The incoming proton stays intact or becomes a low--mass
state, so that there is no colour flow from the proton to the
other final state particles.
 
Similarly to ordinary hadrons, mesons and baryons, the incoming
pomeron is supposed to have a quark and/or a gluon structure, which is
probed through electroweak or strong hard--scattering processes and
expressed by a pomeron structure function. Models based on different
assumptions concerning this quark--gluon structure have been
used for calculations of $p\overline{p}$ and $ep$ scattering processes
(see, e.g., \cite{Ing,Str}). Based on these models, Bruni and Ingelman have
developed a Monte Carlo program \cite{BrIn} to study the event characteristics,
which is already widely used by the HERA collaborations to interpret their
large--rapidity--gap data. If this interpretation is valid, a major task is to
disentangle the quark and gluon distributions of the pomeron. As is the
case in usual hard--scattering processes, one process is not sufficient to
isolate the various parton densities. Several processes must be
considered simultaneously. Deep--inelastic diffractive $ep$
scattering primarily measures the quark distribution of the pomeron, while
the gluon structure function appears as a higher--order QCD correction.
Photoproduction of jets, however,
is sensitive to both the quark and gluon densities already in the
leading order (LO) of QCD. In LO, photoproduction has two
components: a direct one and a resolved one. The latter mechanism has
similar features as jet production in diffractive hadron--hadron scattering.
The hadron structure function is just replaced by the photon structure
function. All these processes can be studied in detail with the Monte Carlo
program cited above.
 
Recently, a number of authors advocated the notion of a direct pomeron
coupling to quarks and gluons \cite{DoLaA,CFS,BeSo}. Donnachie and Landshoff
developed a model in which two gluons emitted coherently from the proton
couple to the quark--antiquark system produced by the incoming real or
virtual photon \cite{DoLaA}. The quark and antiquark thus give rise to two
jets associated with the hard scattering. Such configurations give a
delta function in the pomeron structure function in the same way as direct
photoproduction is equivalent to a delta--function term in the
photon structure function of resolved photoproduction. Since in the
Donnachie--Landshoff model the second gluon of the pomeron is attached directly
to the hard scattering, or to an outgoing parton, the result is higher twist,
i.e., considered as a function of $p_T$, it falls off more strongly than
in the case of
a factorizable direct pomeron coupling. Subsequently, several groups
\cite{CFS,BeSo} suggested other mechanisms which break the usual parton--model
factorization and are effectively proportional to a delta--function term.
These mechanisms, called {\it coherent hard diffraction} \cite{Col} or
{\it lossless diffractive jet production} \cite{BeSo}, have the property
that the total incoming longitudinal momentum of the pomeron is
delivered to the jet system. Both mechanisms produce leading--twist
contributions and are not present in deep--inelastic electron--proton
scattering, i.e., in quasi--real photoproduction, which we are interested
in here primarily, they would occur only in the resolved process.
In addition, it is said explicitly that the CFS mechanism \cite{CFS}
leads to a delta--function term only in the gluon density of the pomeron.
 
So, these mechanisms for creating a quasi--direct pomeron coupling
are very special and their validity is very difficult to assess.
They are just conjectures and not proven consequences of QCD. Under
these circumstances, it is a challenge to look at experimental data
in the diffractive region and to investigate whether such a direct pomeron
coupling is present. To that end, it is necessary to establish features
which are characteristic for a direct pomeron coupling
to quarks and/or gluons in the data. This is the purpose of this work.
 
In section 2, we formulate the model and give the relevant formulas for
calculating the cross section for the production of jets in $ep$ scattering
with almost real photons. There, we specify the structure functions of the
photon and the pomeron and write down the expression for the
pomeron flux. If a direct pomeron coupling exists, this has also
consequences for the diffractive contribution of the deep--inelastic
structure function $F_2$, which in retrospect modifies the input for the
photoproduction cross section.
In section 3, we present numerical results for the three scenarios:
(i) resolved pomeron, (ii) resolved and direct pomeron, (iii)
resolved and direct pomeron plus point--like component in the pomeron
structure function.
The last section is reserved for the conclusions and some outlook to
future work.
 
\vspace{0mm}
\mbox{}
\section{Model for diffractive jet production}
\mbox{}
 
The cross section for the diffractive $ep$ scattering process, depicted in
Fig.~\ref{dia}, is obtained from the following parton model formula:
\bea
\label{d3s}
&&  \hspace{-10mm}
    \frac{d^3\sigma}{dy\,d^2p_T}(ep\rightarrow jet+X) \nonumber\\
&=&
    \sum_{a,b,c,d}\int dx_\gamma \,G_{\gamma/e}(x_\gamma)
                  \int dx_a \,G_{a/\gamma}(x_a)
                  \int dx_{\IP} \,G_{\IP/p}(x_{\IP})
                  \int dx_b \,G_{b/\IP}(x_b)             \nonumber\\
&&
    \times
    \frac{\hat{s}}{\pi} \frac{d\sigma}{d\hat{t}}(a+b\rightarrow c+d)
    \,\delta(\hat{s}+\hat{t}+\hat{u})  \quad,
\eea
where $y$ and $p_T$ are the rapidity and transverse momentum of the jet,
respectively.
$\hat{s}=(p_a+p_b)^2$, $\hat{t}=(p_a-p_c)^2$ and $\hat{u}=(p_b-p_c)^2$
are the familiar Mandelstam
variables for the partonic 2--2 scattering process $a+b\rightarrow c+d$.
$G_{\gamma/e}(x_\gamma)$
stands for the usual Weizs\"acker--Williams formula \cite{WWA}, which
gives the flux of the quasi--real photons. The exact form will be specified
later. $G_{\IP/p}(x_{\IP})$ is an integral over the pomeron flux factor
$f_{\IP/p}(x_{\IP},t)$,
\beq
\label{IPflux}
    G_{\IP/p}(x_{\IP}) = \int_{t_1}^{t_2} dt\,f_{\IP/p}(x_{\IP},t) \quad,
\eeq
where $t$ is the momentum transfer to the proton line.
$t_2=-m_p^2 x_\IP^2/(1-x_\IP)$, with $m_p$ being the proton mass,
is the kinematical upper boundary, while $t_1$ is determined by the experimental
conditions. For $f_{\IP/p}(x_{\IP},t)$ we use the Ingelman--Schlein ansatz
\cite{IS,BrIn,InPr},
\beq
\label{flux}
    f_{\IP/p}(x_{\IP},t)
    = \frac{d^2\sigma/dx_{\IP}\,dt}{\sigma_{\IP\,p\rightarrow X}}
    = \frac{1}{\kappa x_{\IP}}
      \left(a\,e^{\alpha t} + b \,e^{\beta t}\right) \quad,
\eeq
with the parameters $\kappa=2.3\gev^2$, $a=6.38$, $\alpha=8.0\gev^{-2}$,
$b=0.424$ and $\beta=3.0\gev^{-2}$, which were obtained by fitting to the data
of the diffractive cross section and the pomeron--proton total cross section.
Other functional forms of $f_{\IP/p}(x_{\IP},t)$ are given in \cite{DoLaB,Ber};
they are numerically equivalent in the region of small $|t|$, which
dominates the diffractive cross section.
 
For the unknown pomeron structure functions, $G_{b/\IP}(x)$, we make the ansatz
\bea
\label{IPSF}
    G_{u/\IP}(x) \hspace{-2mm}&=&\hspace{-2mm}
                    G_{\overline{u}/\IP}(x) = G_{d/\IP}(x)
                  = G_{\overline{d}/\IP}(x) = \frac{3}{10}(1-x)\quad,
    \nonumber\\
    G_{s/\IP}(x) \hspace{-2mm}&=&\hspace{-2mm}
                    G_{\overline{s}/\IP}(x) = \frac{3}{20}(1-x)\quad,\quad
    G_{c/\IP}(x)  =  G_{\overline{c}/\IP}(x) = 0 \quad,        \nonumber\\
    G_{g/\IP}(x) \hspace{-2mm}&=&\hspace{-2mm}
                    \frac{9}{2}(1-x) \quad.
\eea
The gluon component fulfills
\beq
    \int_0^1 dx\,x\, G_{g/\IP}(x) = \frac{3}{4} \quad,
\eeq
and the quark components have the normalization
\beq
    \sum_{q,\overline{q}} \int_0^1 dx\,x\, G_{q/\IP}(x) = \frac{1}{4} \quad.
\eeq
The sum of the gluon and quark parts obeys the sum rule
\beq
    \int_0^1 dx\,x\, \left(G_{g/\IP}(x)+\sum_{q,\overline{q}}
    G_{q/\IP}(x)\right) = 1 \quad,
\eeq
which is usually assumed also by other authors \cite{BrIn}.
Since the exchanged pomeron is not a physical particle, there is no
rigid justification for this sum rule, although some motivation can be given.
 
The quark content of the pomeron can be measured in diffractive deep--inelastic
$ep$ scattering (DDIS). Preliminary data from the H1 and ZEUS Collaborations
exist, which will be used for comparisons later. The gluon content of the
pomeron contributes to DDIS through the order $\alpha_s$ process $\gamma g
\rightarrow q\overline{q}$ and is therefore suppressed. This means that,
in a first approximation, DDIS is determined by the quark and antiquark
components of the pomeron structure function. The gluon component enters when
the pomeron structure function is evolved to larger $Q^2$, which, however,
we shall not perform in the crude study presented here.
 
For the calculation of the photoproduction of jets, we need the hard--scattering
cross sections $d\sigma/d\hat{t}$ in (\ref{d3s}). We consider four components,
DD, DR, RD, RR. The first letter indicates whether the incoming quasi--real
photon interacts directly (D) or as a resolved photon (R) with the
incoming quarks and gluons. The second letter characterizes
the pomeron interaction; D stands for the direct coupling of
the pomeron and R for a resolved pomeron described by the
structure functions in (\ref{IPSF}). The resolved--pomeron
hard--scattering cross sections are the usual ones. In the
case of a direct photon coupling, we have the photon--gluon--fusion
($\gamma g \rightarrow q\overline{q}$) cross section, where a gluon
inside the pomeron is struck, and the gluon--Compton
($\gamma q \rightarrow g q$) cross section, where a quark or antiquark
is pulled out of the pomeron. For a resolved photon and a resolved
pomeron (RR), we have the well known 2--2
quark and gluon scattering cross sections.
 
For the calculation of the cross sections with a direct pomeron coupling, we
restrict ourselves to a direct coupling with quarks (antiquarks) of the form
\beq
\label{Lint}
    {\cal{L}}_{int} = c\,\overline{q}(x) \gamma_\mu q(x)\,\phi^\mu (x) \quad.
\eeq
Here $c$ is the coupling constant. In (\ref{Lint}), we assumed that the
pomeron, $\phi$, couples to quarks like a vector particle. This contradicts the
fact that the pomeron behaves like a $C=+1$ exchange, whereas (\ref{Lint})
transforms invariantly under $C$ only if $\phi$ has $C=-1$. It is an old
observation, however, that, in soft processes like diffractive scattering,
the pomeron couples to quarks effectively
rather like an isoscalar photon, i.e., with a constant $\gamma_\mu$
coupling but with a Regge signature factor, which endows it with even $C$
parity \cite{LaPo,JaLa}. With this ansatz, Donnachie and Landshoff were able to
give a realistic description of all high--energy elastic and diffractive cross
sections applying the additive quark model and including less dominant Regge
exchanges \cite{DoLaB}. Pomeron exchange with the coupling of (\ref{Lint}) also
correctly predicts the approximate helicity conservation that is observed
experimentally.
 
With such a coupling, we obtain the following hard--scattering cross section of
$\gamma \IP \rightarrow q\overline{q}$, written in terms of the invariants,
$\hat{s}$, $\hat{t}$, $\hat{u}$, introduced earlier:
\beq
\label{dsdt}
    \frac{d\sigma}{d\hat{t}}\left(\gamma \IP \rightarrow q\overline{q}\right)
 = \frac{1}{16\pi(\hat{s}-t)^2} \,6e^2e_q^2 c^2
    \left(\frac{\hat{u}}{\hat{t}} + \frac{\hat{t}}{\hat{u}}
          + \frac{2\hat{s}t}{\hat{u}\hat{t}}\right) \quad,
\eeq
where $e_q$ is the electric charge of $q$ in units of the positron charge, $e$.
Here, $t$ is the momentum transfer at the proton vertex: $t=m_{\IP}^2$,
if the pomeron is timelike with mass $m_{\IP}$, while $t<0$ in our application.
Of course, (\ref{dsdt}) is identical to the cross section of
$\gamma\gamma^* \rightarrow q\overline{q}$, where $\gamma^*$ is an
off--shell photon with momentum squared $t$.
 
Introducing the transverse momentum $p_T$ and the rapidity $y$ of the
outgoing jet, we have
\beq
    \cosh y = \frac{\sqrt{\hat{s}}}{2p_T} \quad,
    \quad \hat{u}\hat{t} = \frac{p_T^2}{\hat{s}}\left(\hat{s}-t\right)^2 \quad,
    \quad \hat{u} + \hat{t} = -\hat{s} + t \quad,
\eeq
so that
\beq
\label{dsdp}
    \frac{d\sigma}{dp_T^2}\left(\gamma \IP \rightarrow q\overline{q}\right)
 = \frac{1}{16\pi\hat{s}(\hat{s}-t)}
    \frac{1}{\sqrt{1-4p_T^2/\hat{s}}} \,6e^2e_q^2 c^2
    \left[\frac{\hat{s}^2+\hat{t}^2}{(\hat{s}-t)^2} \frac{\hat{s}}{p_T^2}
    -2\right] \quad.
\eeq
For $\hat{s} \gg p_T^2$, the cross section decreases only like $1/p_T^2$, i.e.,
it exhibits a rather mild $p_T$ dependence. Furthermore, the
rapidity is completely fixed by $p_T$ and $\hat{s}$. Under the assumption that
the pomeron couples like a scalar particle (this would not allow the
description of elastic and diffractive scattering proposed in \cite{DoLaB}),
(\ref{dsdp}) would only change slightly to
\beq
    \frac{d\sigma}{dp_T^2}\left(\gamma \IP \rightarrow q\overline{q}\right)
 = \frac{1}{16\pi\hat{s}(\hat{s}-t)}
    \frac{1}{\sqrt{1-4p_T^2/\hat{s}}} \,6e^2e_q^2 c^2
    \frac{\hat{s}^2+\hat{t}^2}{(\hat{s}-t)^2} \frac{\hat{s}}{p_T^2}
    \quad,
\eeq
i.e., for $p_T^2 \ll \hat{s}$ the result coincides with the one for the vector
coupling.
In the following, we shall assume that the direct pomeron coupling is
characterized by (\ref{Lint}), i.e.,
we shall employ (\ref{dsdp}) for the hard--collision cross section.
 
In our model for the direct pomeron coupling, the pomeron behaves like a real
photon target for $t=0$. Similarly to $\gamma\gamma$ scattering, the
$\gamma \IP \rightarrow q\overline{q}$ cross section also contributes to the
pomeron structure function, i.e., to deep--inelastic $ep$ scattering. Then the
incoming photon in (\ref{dsdp}) is off--shell with invariant mass squared
$(-Q^2)$. In inelastic $e\gamma$ scattering, this represents the contribution of
the {\it point--like} photon in the photon structure function. Similarly,
we have a {\it point--like} contribution to the quark part of the pomeron
structure function, which is obtained from (\ref{dsdp}) by integrating over
$p_T^2$. In the form of (\ref{dsdp}), the cross section has a collinear
singularity at $p_T=0$. To remove it, we render the quark masses, $m_q$,
nonvanishing as usual \cite{GoSt} and obtain
\bea
\label{poi}
    G_{q/\IP}^{pl}(x,Q^2) &=& \frac{N_c}{8\pi^2}\,c^2 \Bigg\{
    \beta \bigg[-1+8x\left(1-x\right)-\frac{4m_q^2}{Q^2}x\left(1-x\right)\bigg]
    \nonumber\\
&&
    + \bigg[x^2+\left(1-x\right)^2+\frac{4m_q^2}{Q^2}x\left(1-3x\right)
            -\frac{8m_q^4}{Q^4}x^2 \bigg]
    \ln\frac{1+\beta}{1-\beta}\Bigg\}    \quad ,
\eea
with
\bea
    \beta &=& \sqrt{1-\frac{4m_q^2x}{Q^2(1-x)}} \quad.
    \nonumber
\eea
The quark masses are chosen as $m_u=m_d=0.3\gev$, $m_s=0.5\gev$ (i.e.,
equal to the constituent masses) and $m_c=1.5\gev$. $Q^2<0$ is the
momentum squared of the off--shell incoming photon. In a more
realistic treatment, the mass parameter is
replaced by $\Lambda_{QCD}$. It is known that (\ref{poi}) is a reasonable
approximation for the case of the point--like part of the photon structure
function. For large $Q^2$, we have
\beq
\label{poia}
    G_{q/\IP}^{pl}(x,Q^2) = \frac{N_c}{8\pi^2}\,c^2
    \bigg[x^2+\left(1-x\right)^2\bigg] \ln \frac{Q^2(1-x)}{m_q^2x} \quad.
\eeq
 
As expected, the point--like part of the quark distribution of the
pomeron dominates at large $Q^2$. This result is unusual for the structure of
such a complicated object as the pomeron. As a first guess, one would expect
that the quark distribution function of the pomeron would behave more like the
one of a meson, a $f$ or a $\rho^o$ meson say, which is
supposed to have no point--like component. Whether this is true can be tested by
inspecting the data and trying to eventually place an upper limit on the
coupling $c$.
Furthermore, the $x$ dependence of $G_{q/\IP}^{pl}(x,Q^2)$ is quite
different from the behaviour of the corresponding structure function of a
resolved pomeron in (\ref{IPSF}).
 
To LO in $\alpha_s$, only the quark distribution of the
pomeron enters into the deep--inelastic $e\IP$ structure function
$F_2^{\IP}(x,Q^2)$, which is
\beq
\label{F2IP}
    F_2^{\IP}(x,Q^2) = \sum_q e_q^2\,x\,\left[G_{q/\IP}(x)
    + G_{\overline{q}/\IP}(x) + 2\,G_{q/\IP}^{pl}(x,Q^2)\right] \quad.
\eeq
In Fig.~\ref{plikec},
we have plotted the point--like distribution function for the three
quark masses $m_q=0.3$, $0.5$, $1.5\gev$ and the four values $Q^2=2.5^2$,
$5.0^2$, $7.5^2$, $10.0^2\gev^2$. Notice the logarithmic singularity
at $x=0$. Apart from that, $G_{q/\IP}^{pl}(x,Q^2)$ falls off with $x$
increasing and increases with $Q^2$ even at large $x$, in contrast to what one
expects for a hadron--like object.
For calculating $G_{q/\IP}^{pl}$, we have chosen $c=1$.
This value is somewhat smaller than the one deduced from fits to total
cross sections and elastic--scattering data by Donnachie and Landshoff
\cite{DoLaB,JaLa}. Figure~\ref{plikec}d shows the total contribution
of the point--like part to $F_2^\IP(x,Q^2)/(2x)$ for the same $Q^2$ values,
i.e., the sum
of the $u$, $d$, $s$, $c$ contributions with the appropriate charge factors.
The dents in the curves are caused by the charm threshold.
 
Under the assumption that factorization is applicable,
$F_2^\IP(x,Q^2)$ can be used to calculate the diffractive contribution
to the deep--inelastic structure function $F_2^{diff}(x,Q^2)$ of the proton.
The relation is
\beq
\label{F2diff}
    F_2^{diff}(x,Q^2) = \int_x^{x_0} dx_\IP\,\int_{t_1}^{t_2} dt\,
    f_{\IP/p}(x_\IP,t) F_2^\IP(x/x_\IP,Q^2) \quad.
\eeq
Here $f_{\IP/p}(x_\IP,t)$ is the pomeron flux factor given by (\ref{flux}).
The diffractive contribution is concentrated at small $x$, and
we shall mostly take $x_0=0.01$ as in the experimental analyses by the
H1 and ZEUS Collaborations. $t_2$ is defined below (\ref{IPflux}), and we
choose $t_1=-1.0\gev^2$. The results for $F_2^{diff}(x,Q^2)$ at
$Q^2=8.5$, $15$, $30$, $60\gev^2$ are compared to preliminary
H1 data \cite{Fel} in Fig.~\ref{F2df}.
We see that the contribution without direct pomeron
($c=0$) is too small. Of course, this can be changed by increasing
$G_{q/\IP}(x)$ in relation to $G_{g/\IP}(x)$ in (\ref{IPSF}); in its present
form, the relation is 1:3.
With our choice of (\ref{IPSF}), the agreement with the data is improved
significantly when we include the pointlike contribution with $c=1$.
Clearly, Fig.~\ref{F2df} only demonstrates that $c=1$ is
a consistent value for the direct pomeron coupling when the resolved pomeron
is described by (\ref{IPSF}). That our curves for $c=1$ lie somewhat above
the data points can be tolerated, since the data are
obtained with the cut $\eta_{max} \le 1.5$, i.e., only a fraction, presumably
not more than 50--70\%, of the diffractive contribution is included in the
data points. As expected from the cut on the invariant mass $M_X$, i.e.,
$x_0=0.01$, the data for the diffractive part of $F_2$ are nonzero only for
$x<10^{-2}$. In the case of $F_2^{diff}(x,Q^2)$, we could, in principle,
accommodate a larger value of $c$ by correspondingly scaling down
$G_{q/\IP}(x)$ relative to $G_{g/\IP}(x)$.
By fitting simultaneously $F_2^{diff}(x,Q^2)$ and the jet--photoproduction
cross sections to the data, it should be possible to place an upper bound
on $c$.
 
We emphasize that the proton momentum transfer is neglected in (\ref{poi})
and hence in (\ref{F2IP}) and in the $e\IP$ structure function in
(\ref{F2diff}). This is justified, since in (\ref{F2diff})
the integral over $t$ is dominated by the minimum $|t|$, which is
$|t_2|\approx 0$. Of course, it would be worthwhile to consider
$F_2^{diff}(x,Q^2)$ in (\ref{F2diff}) for fixed $t$ and to study the
full $t$ dependence.
 
We now turn to the jet cross sections of diffractive photoproduction.
The calculation of the one-- and two--jet inclusive cross section is
straightforward and proceeds from (\ref{d3s}). In this formula, the integration
limits must be specified in terms of the external variables. For the one--jet
inclusive cross section, they are $y$, $p_T$ and $S=(p_e+p_p)^2$,
the energy squared of the $ep$ centre-of-mass (c.m.) system. For the case of
RR, i.e., resolved $\gamma$ and resolved $\IP$, the incoming parton momenta are
$p_a=x_\gamma x_a p_e$ and $p_b=x_\IP x_b p_p$, so that $\hat{s}=x_\gamma
x_a x_\IP x_b S$, $\hat{t}=-x_\gamma x_a \sqrt{S}\,p_T e^{-y}$ and
$\hat{u}=-x_\IP x_b \sqrt{S}\,p_T e^{y}$ (see Fig.~\ref{dia}).
It is clear that all momentum
fractions are limited in general to the interval $[0,1]$.
In our formulas, we have chosen the $ep$ c.m.\ system, and the momentum of the
incoming electron is taken along the positive $z$ direction. The transformation
to the HERA system will be done when we present the numerical results
in the next section. From phase space, we have as the integration
limits in (\ref{d3s})
\bea
    x_\gamma^{min} \hspace{-2mm}&=&\hspace{-2mm}
    \frac{e^y}{\frac{\sqrt{S}}{p_T} - e^{-y}} \quad, \quad
    x_a^{min} = \frac{e^y}{x_\gamma(\frac{\sqrt{S}}{p_T}-
    \frac{e^{-y}}{x_\IPs})} \quad,
    \nonumber\\
    x_\IP^{min} \hspace{-2mm}&=&\hspace{-2mm}
    \frac{e^{-y}}{\frac{\sqrt{S}}{p_T} - \frac{e^{y}}{x_\gamma}} \quad, \quad
    x_b^{min} = \frac{e^{-y}}{x_\IP(\frac{\sqrt{S}}{p_T}-
    \frac{e^{y}}{x_\gamma x_a})} \quad.
\eea
 
As usual, the functions $G_{i/j}(x_i)$ stand for the probability to find a
parton $i$ with momentum fraction $x_i$ in parton $j$, where $i$ and $j$
may be also the electron, photon or pomeron. $G_{\gamma/e}(x)$ is described
by an improved Weizs\"acker--Williams function \cite{RoSo}
\beq
    G_{\gamma/e}(x) = \frac{\alpha}{2\pi}\left[\frac{1+(1-x)^2}{x}
    \ln\frac{Q_{max}^2}{Q_{min}^2} - \frac{2(1-x)}{x}\right] \quad,
\eeq
where $Q_{max}^2=0.01\gev^2$ and $Q_{min}^2=m_e^2 x^2/(1-x)$.
We adopt the photon structure functions from \cite{GRV}.
The contribution due to a direct photon and a resolved pomeron (DR)
is calculated from (\ref{d3s}) using $G_{a/\gamma}(x_a)=\delta(1-x_a)$ and
the hard--scattering cross sections of $\gamma q \rightarrow g q$
and $\gamma g \rightarrow q\overline{q}$.
 
Experimentally, the rapidities are limited by $\eta_{max}\le 1.5$.
In the one--jet inclusive cross section, the integration over the
rapidity of the second jet goes beyond this limit. To incorporate
the limit, we need to consider the two--jet cross section. It is
obtained from
\bea
\label{d3s2}
&&  \hspace{-10mm}
    \frac{d^3\sigma}{dy_c\,dy_d\,dp_T^2}(ep \rightarrow j_c\,j_d+X)  \\
&=&
    \sum_{a,b,c,d} \int_{x_\gamma^{min}}^{x_\gamma^{max}}
    dx_\gamma\,G_{\gamma/e}(x_\gamma)
    \int_{x_\IPs^{min}}^{x_\IPs^{max}} dx_\IP\,G_{\IP/p}(x_\IP)\,
    x_a\,G_{a/\gamma}(x_a)\,x_b\,G_{b/\IP}(x_b)
    \frac{d\sigma}{d\hat{t}}(ab\rightarrow cd) \quad, \nonumber
\eea
where
\bea
    x_a &=& \frac{p_T}{x_\gamma \sqrt{S}}\left(e^{y_c}+e^{y_d}\right) \quad,
    \nonumber\\
    x_b &=& \frac{p_T}{x_\IP\sqrt{S}}\left(e^{-y_c}+e^{-y_d}\right) \quad.
\eea
So, for fixed $x_\gamma$ and $x_\IP$, one can directly extract information
on the structure functions of the photon and the pomeron by varying
$p_T$, $y_c$ and $y_d$. The lower limits of the $x_\gamma$ and $x_\IP$
integrals are
\bea
    x_\gamma^{min} &=& \frac{p_T}{\sqrt{S}}\left(e^{y_c}+e^{y_d}\right) \quad,
    \nonumber\\
    x_\IP^{min} &=& \frac{p_T}{\sqrt{S}}\left(e^{-y_c}+e^{-y_d}\right) \quad.
\eea
Clearly, $p_T$, $y_c$ and $y_d$ are limited by the conditions
$x_\gamma^{min} \leq x_\gamma^{max}\leq 1$ and
$x_\IP^{min} \leq x_\IP^{max}\leq 1$.
The rapidities $y_c$ and $y_d$ are defined in such a way that the
incoming electron travels in the positive $z$ direction, i.e.,
$p_c=p_T(\cosh y_c,1,0,\sinh y_c)$ and $p_d=p_T(\cosh y_d,-1,0,\sinh y_d)$.
In the numerical discussion, we shall flip the sign of the rapidity so as
to be in conformity with HERA standards.
The one--jet inclusive cross section is obtained by identifying
$y \equiv y_c$ and integrating (\ref{d3s2})
over $y_d\in [y_{d_1},y_{d_2}]$, where
$y_{d_1}=-\ln(\sqrt{S}/p_T-e^{-y_c})$
and $y_{d_2}=\ln(\sqrt{S}/p_T-e^{y_c})$.
In our case, the two--jet cross section is
the integral over $y_d$ with the lower limit
$y_{d_1}'=\max(y_c,y_{d_1})$, so that the rapidity gap is always determined by
$y=y_c$. In the following, this cross section is denoted by
$d^2\sigma(2 jet)/dy\,dp_T^2$, while the
inclusive one--jet cross section is called
$d^2\sigma(1 jet)/dy\,dp_T^2$.
 
As the last point, we write down the cross section for the DD process,
with direct photon and direct pomeron in the initial state. This amounts to
putting $x_a = x_b = 1$ in (\ref{d3s}), so that the inclusive
one--jet cross section is
\beq
\label{ddone}
\frac{d^2\sigma}{dy\,dp_T^2}(1 jet)
=\sum_{c,d}
   \int_{x_\gamma^{min}}^{x_\gamma^{max}} dx_\gamma\,G_{\gamma/e}(x_\gamma)\,
   x_\IP^2\frac{\sqrt{S}}{p_T}\, e^y \,G_{\IP/p}(x_\IP)
   \frac{d\sigma}{d\hat{t}}(\gamma\IP \rightarrow cd) \quad,
\eeq
where
\beq
    x_\IP = \frac{e^{-y}}{\frac{\sqrt{S}}{p_T}
    -\frac{e^y}{x_\gamma}}     \quad,
\eeq
i.e., $x_\IP$ is fixed by $y$, $p_T$ and $x_\gamma$. On the other hand,
the condition $x_\IP\leq x_\IP^{max}$ determines the lower bound of
integration in (\ref{ddone}),
\bea
    x_\gamma^{min} &=& \frac{e^{y}}{\frac{\sqrt{S}}{p_T}
    -\frac{e^{-y}}{x_\IPs^{max}}}     \quad.
\eea
In the numerical analysis, we shall exclude $x_\gamma$ values outside the
interval $[0.3,0.7]$.
Then, we shall have $0.3 \leq x_\gamma^{min} \leq x_\gamma^{max} = 0.7$.
For fixed $p_T$, the allowed rapidity interval is fully determined by the
condition $x_\gamma^{min} \leq x_\gamma \leq x_\gamma^{max}$.

\section{Results}
\mbox{}
 
In this section, we present results for the one-- and two--jet cross sections as
defined in the last section with kinematical constraints as used in the
data analyses by H1 and ZEUS. The results are given for the HERA
frame with $E_e=26.7\gev$, $E_p=820\gev$ and the positive $z$ axis pointing in
the incoming--proton direction.
We choose $x_0=0.01$ for the upper cut on $x_\IP$
and use the structure functions for electron, photon and pomeron
that were specified in the previous section. Figure \ref{yds}
shows $d^2\sigma/dy\, dp_T^2$
for the one--jet inclusive (a--c) and for the two--jet cross section (d--f).
The cross sections are plotted as a function of rapidity at $p_T=5\gev$.
We shall first concentrate on one--jet inclusive production.
In Fig.~\ref{yds}a, the direct pomeron coupling, $c$, is put to zero.
Therefore, we have only results for DR and RR.
In Fig.~\ref{yds}a, we give also the one--jet inclusive cross section for
the usual photoproduction of jets for comparison. As is well known, at
$p_T=5\gev$, the resolved--photon cross section (R) dominates the
direct--photon cross section (D).
The diffractive jet cross sections are limited to rapidity values
$y\simlt 1$. (The precise interval will be specified below.)
This is due to the $x_0$ cut or, experimentally, the cut on
$M_X$, the invariant mass of the diffractively produced final state.
The DR and RR components have cross sections of the same
order. The DR cross section exceeds the RR one at negative rapidities.
In Fig.~\ref{yds}b, the direct pomeron coupling is added with $c=1$.
This produces a significant cross section in DD, i.e., with
direct photon and direct pomeron coupling. This cross section
is peaked for negative rapidities and
is of similar magnitude as the DR component. The RD component is small compared
to all others. We emphasize that, at $p_T=5\gev$, the direct pomeron
predominantly contributes together with the {\it direct photon}.
 
Figure~\ref{yds}c displays the inclusive one--jet cross sections including
also the
point--like part of the quark distribution of the pomeron given by
(\ref{poi}), again with the normalization $c=1$.
This affects only the DR and RR components.
By comparing the respective curves in Figs.~\ref{yds}b, c,
we see that the point--like piece of the pomeron structure function leads to
a modest increase of these cross sections, by some $10\%$.
This moderate increase is explained by the fact that the gluon structure
function of the pomeron is the dominant part in the DR and RR components.
In Figs.~\ref{yds}d--f, we repeat the analyses of Figs.~\ref{yds}a--c
for the two--jet cross sections.
The restriction on the rapidity of the second jet leads to a decrease of the
cross sections and limits the rapidity range in the electron direction.
In the c.m.\ frame, the lower edge of the rapidity spectrum now appears at
$-\ln\left(x_\gamma^{max}\sqrt S/2p_T\right)$, while
in the one--jet case it occurs at
\beq
-\ln\left[{x_\gamma^{max}\over2}\left({\sqrt S\over p_T}
+\sqrt{{S\over p_T^2}-{4\over x_\gamma^{max}x_\IP^{max}}}\,\right)\right]
\approx-\ln{x_\gamma^{max}\sqrt S\over p_T}\quad.\nonumber
\eeq
In both cases, the upper edge lies at
\beq
\ln\left[{x_\IP^{max}\over2}\left({\sqrt S\over p_T}
+\sqrt{{S\over p_T^2}-{4\over x_\gamma^{max}x_\IP^{max}}}\,\right)\right]
\approx\ln{x_\IP^{max}\sqrt S\over p_T}\quad.\nonumber
\eeq
The rapidity in the laboratory frame is shifted by $+(1/2)\ln(E_p/E_e)$
relative to the c.m.\ rapidity.
>From these results it is obvious that the rapidity distribution for
a fixed $p_T$ value near $5\gev$ is not useful to distinguish
the direct-- and resolved--pomeron contributions.
The two combinations DD $+$ RD and DR $+$ RR are
very similar in their rapidity dependence. Figures~\ref{sumyds}a, b
show the sum over the various components in Fig.~\ref{yds} for
the inclusive one--jet and two--jet cross sections, respectively.
In Fig.~\ref{sumyds}a, we compare our results with recent H1 data \cite{Fel}.
Since these data are given
as event rates, we can only compare the shapes. The data point with
lowest rapidity is fitted to the theoretical curve, which is in agreement
with the other data points within the experimental errors.
For simplicity, we choose the normalization factor to be 1/2.
 
The $p_T$ dependence of the cross sections for fixed
rapidity is more discriminative. This can be seen from
the results in Fig.~\ref{pts}. In Figs.~\ref{pts}a--c,
we have plotted the inclusive one--jet
cross section as a function of $p_T$ for fixed $y=0$, again for the three
cases $c=0$, $c=1$ and $c=1$ with point--like pomeron structure function.
The DD component has a much slower fall--off in $p_T$ as compared to the
DR component and dominates the other components for $p_T \simgt 5\gev$.
This harder $p_T$ dependence is expected, since the hard--scattering cross
section of $\gamma \IP \rightarrow q\overline{q}$  is proportional to
$1/p_T^2$. In the DR component, this
dependence is softened through the pomeron structure functions. The two--jet
cross sections show an almost identical pattern, since the additional
restriction for the second jet is less effective, if $y$ is held fixed
near the maximum of the rapidity distribution.
 
The shallow $p_T$ dependence of the DD (and, to a minor extent, of the RD)
component
is still visible in the sum over all components. This is seen in
Figs.~\ref{sumpts}a, b for the inclusive one--jet and two--jet cross
sections, respectively, when we compare the curves for $c=1$ and $c=0$.
In Figs.~\ref{sumpts}a, b, we also compare our results with recent H1
\cite{Fel} and ZEUS data \cite{ZEUSC}, respectively. Similarly to
Fig.~\ref{sumyds}a, we can only compare the shapes.
The points with lowest $p_T$ are adjusted so as to match approximately
with the theoretical curves, again using a normalization factor of 1/2.
The data seem to show the shallow $p_T$ dependence characteristic for the
combined direct--pomeron contribution, DD $+$ RD.
However, this observation should be taken with a grain of salt.
Firstly, these data are uncorrected. Secondly, for a serious
comparison, our model should include hadronization, which is certainly
important for the low effective c.m.\ energies in the $\gamma\IP$ channel.
Therefore, we believe that it is premature to draw any firm conclusions from
these comparisons.
 
There are two other features in the ZEUS data that are indicative of a
direct pomeron coupling. The $x_\gamma\equiv x_a$ distribution is peaked near
$x_\gamma =1$, and the $x_b$ distribution, i.e., the dependence on the
pomeron--structure--function variable $x_b$, has also a maximum near
$x_b=1$.
But also here, conclusions can be drawn only when corrected data are available
and hadronization effects are included in our model.

\section{Conclusions and Outlook}
\mbox{}
 
In this work, we studied the effect of a direct pomeron coupling
to quarks, which leads to lossless diffractive jet production.
We considered $ep$ scattering under HERA conditions with almost real photons.
We tried to conform to the experimental conditions of the H1 and ZEUS
experiments as much as possible. Only LO processes were investigated, and the
hadronization of the parton jets was neglected. The most important signature
of a direct pomeron coupling is the $p_T$ distribution in inclusive
one--jet and two--jet production. It leads to a much flatter $p_T$
distribution than in the case of a resolved pomeron. The direct photon coupling
gives the dominant contribution at $p_T \simgt5\gev$. Most of our results
are model--dependent, since we have no a--priori knowledge about the strength of
the pomeron coupling and about the relation of the quark to the gluon
components of the pomeron structure function. We considered the gluon part
to be dominant over the quark part in the ratio $3:1$. So far, the
existing preliminary experimental information on jet production
with almost real photons in rapidity--gap events observed by H1 and ZEUS
is consistent with this assumption. Definite conclusions are not possible
at this time,
since these data are not yet corrected and our model does not contain
hadronization corrections, which are likely to be important at these low
energies in the $\gamma\IP$ c.m.\ system.
 
We are aware of the fact that the assumption of a direct pomeron coupling
to quarks is quite
unconventional, since the pomeron is a very complex object, which, at first
hand, is not expected to behave similarly to a photon.
On the other hand, theory cannot guide us.
So, only data can tell whether our assumptions are realistic.
 
There exist several modifications of the ideas presented in this work.
Firstly, the direct pomeron coupling to quarks could contain an additional form
factor which depends on $p_T$. In this case, we would expect that the $p_T$
dependence of the DD component is not as flat as in Fig.~\ref{pts}.
Therefore, it would be much harder to detect such a direct coupling. Secondly,
the direct coupling of the pomeron might be to gluons and not to quarks---or
to both. For a direct gluon coupling of the pomeron, the direct--photon
contribution does not exist in LO. Therefore, such a coupling could be seen
only in connection with a resolved photon. Of course, for this case, there
is also no point--like component in the pomeron structure function. If
direct pomeron couplings both to quarks and to gluons exist, it will be
difficult to disentangle the two.
 
It is conceivable that, when the experimental range of $x_\IP$ can be increased,
also other Regge exchanges between the primary proton vertex and the
hard--scattering process will become relevant. Such exchanges, like $\pi$,
$\rho$, etc., have been considered in the past as a method for extracting the
deep--inelastic structure of Reggeons \cite{SACS} when the incoming photon is
highly virtual.
Such exchanges are also possible in $ep$ scattering with almost real
photons. In this case, a direct $\pi$ ($\rho$, etc.) coupling to quarks
is even more natural than a direct pomeron coupling to quarks. Actually, in a
recent publication \cite{E683} by the E--683 Collaboration at Fermilab
on jet production by real photons in the fixed--target energy range, it is
reported that the energy flow in the photon (forward) direction is very
similar in $\gamma p$ and $\pi p$ reactions at comparable energies.
Since $\gamma p$ processes at such low c.m.\ energies are dominated by
{\it direct} photoproduction, this result can be interpreted by assuming that
also the pion has an appreciable direct coupling to quarks, in contrast to
na\"\i ve expectations.
 
Similar results were reported earlier by the E--609 Collaboration \cite{Nau}
for hard $\pi p$ collisions at $200\gev$ on a fixed target. These
authors found evidence for two--jet events with little or no energy flow
in the forward direction. They interpreted these results in terms of a
higher--twist process suggested by Berger and Brodsky \cite{BeBr}. In this
model, the pion structure function has a delta--function component with a
form factor that is related to the electromagnetic form factor of the pion.
Except for these higher--twist factors, which reduce the cross section
and produce a steeper $p_T$ dependence, the cross section has the same
characteristics as the one of jet production in $\gamma p$ collisions with a
direct photon.
 
It is possible that the pomeron, too, has such a higher--twist component.
This means that, independently of whether the pomeron model with a
leading--twist direct coupling proposed in this work is true or not,
it might be reasonable to look for lossless diffractive jet production in $ep$
collisions, which could be a signature for a higher--twist direct coupling.
We remark that the c.m.\ energies in the $\gamma p$ collisions considered in
this work are of the same order as those in the $\pi p$ fixed--target
experiments.
 
\newpage
\bigskip
\centerline{\bf ACKNOWLEDGMENTS}
\smallskip\noindent
We would like to thank G.\ Wolf for useful comments on the ZEUS data on
diffractive photoproduction of jets and G.\ Ingelman for carefully reading
the manuscript.
One of us (BAK) is indebted to the KEK Theory Group for the warm hospitality
extended to him during his visit, when this work was finalized.

 
\vskip-6cm
 
\begin{figure}[t]
\centerline{\bf FIGURE CAPTIONS}
\smallskip
\caption{}
\label{dia}
Generic diagram for the diffractive $ep$ scattering
process with a resolved photon $\gamma$, a resolved pomeron $\IP$ and the
hard subprocess $HS$.
 
\caption{}
\label{plikec}
The $x$ dependence of the point--like distribution function,
$G_{q/\IP}^{pl}(x,Q^2)$, for various values of $Q^2$ and for the three
quark masses a) $m_q=0.3\gev$, b) $m_q=0.5\gev$, c) $m_q=1.5\gev$; the
$e_q^2$--weighted sum, $\sum_q e_q^2 G_{q/\IP}^{pl}(x,Q^2)$,
where $q=u,d,s,c$, is shown in d).
 
\caption{}
\label{F2df}
$F_2^{diff}(x,Q^2)$ compared to preliminary H1 data.
The solid curves represent only the contributions from
quarks in the resolved pomeron, while for the dashed lines
the point--like contribution with $c=1$ is included.
 
\caption{}
\label{yds}
The rapidity distributions of the one--jet [a)--c)]
and two--jet [d)--f)] cross sections for fixed $p_T=5\gev$ in the $ep$
laboratory system. Here $y$ is defined to be positive for jets travelling in
the proton direction.
Figures a), d) show the non--diffractive cross
sections with a direct (D) and resolved (R) photon in comparison with the
DR and RR contributions in our model. Due to $c=0$, there are no
DD or RD contributions. In Figs.~b), e), these contributions are included
for $c=1$ but without the point--like (pl) component of the pomeron. This
contribution is switched on finally in Figs.~c), f).
 
\caption{}
\label{sumyds}
The $y$ distributions of the a) one--jet and b) two--jet cross
sections for fixed transverse momentum $p_T=5\gev$ after summing up the various
contributions shown in Fig.\ 4. For a direct comparison with data,
the one--jet events from H1 with suitable normalization are also shown.
\end{figure}
 
\begin{figure}[t]
\caption{}
\label{pts}
The $p_T$ spectra of the one--jet [a)--c)] and two--jet [d)--f)]
cross sections for fixed rapidity $y=0$. Figures a)--f)
have to be considered in full analogy to Figs.~4a)--f).
 
\caption{}
\label{sumpts}
The $p_T$ spectra of the a) one--jet and b) two--jet cross
sections for fixed rapidity $y=0$ after summing up the various
contributions shown in Fig.~\ref{pts}. For a direct comparison with data,
the one--jet events from H1 and the two--jet events from ZEUS
with suitable normalizations are also shown.
Like in Figs.~4a), d), we have also plotted the non--diffractive
result D+R.
\vspace{155mm}
\mbox{}
\end{figure}
 
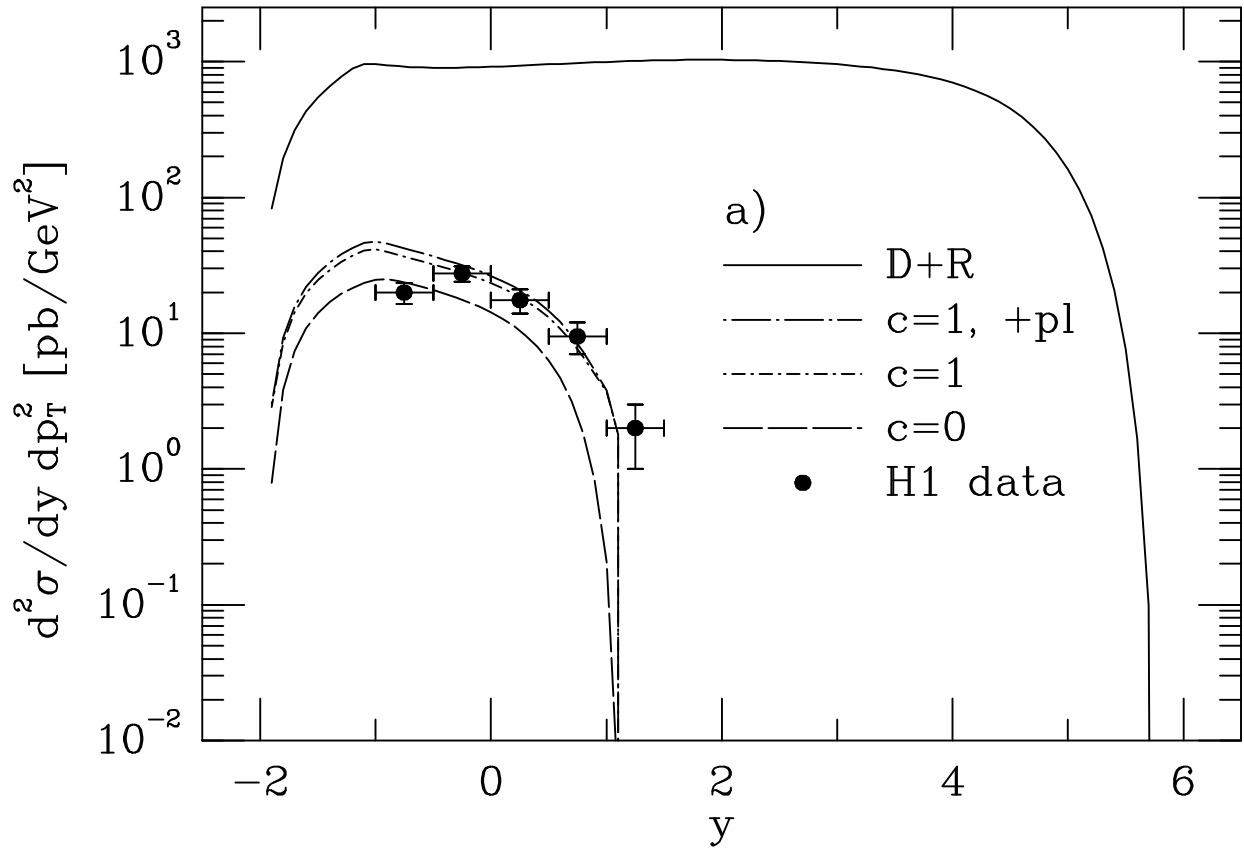
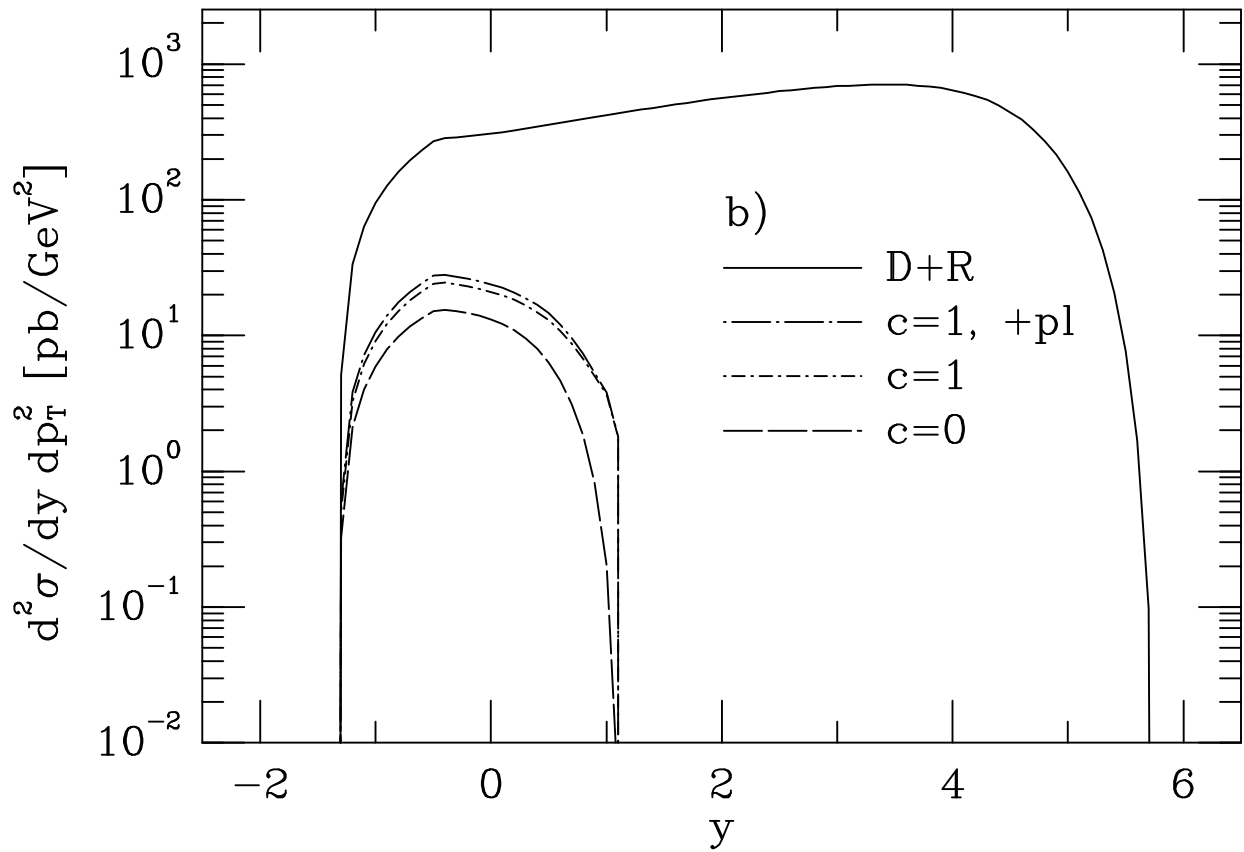
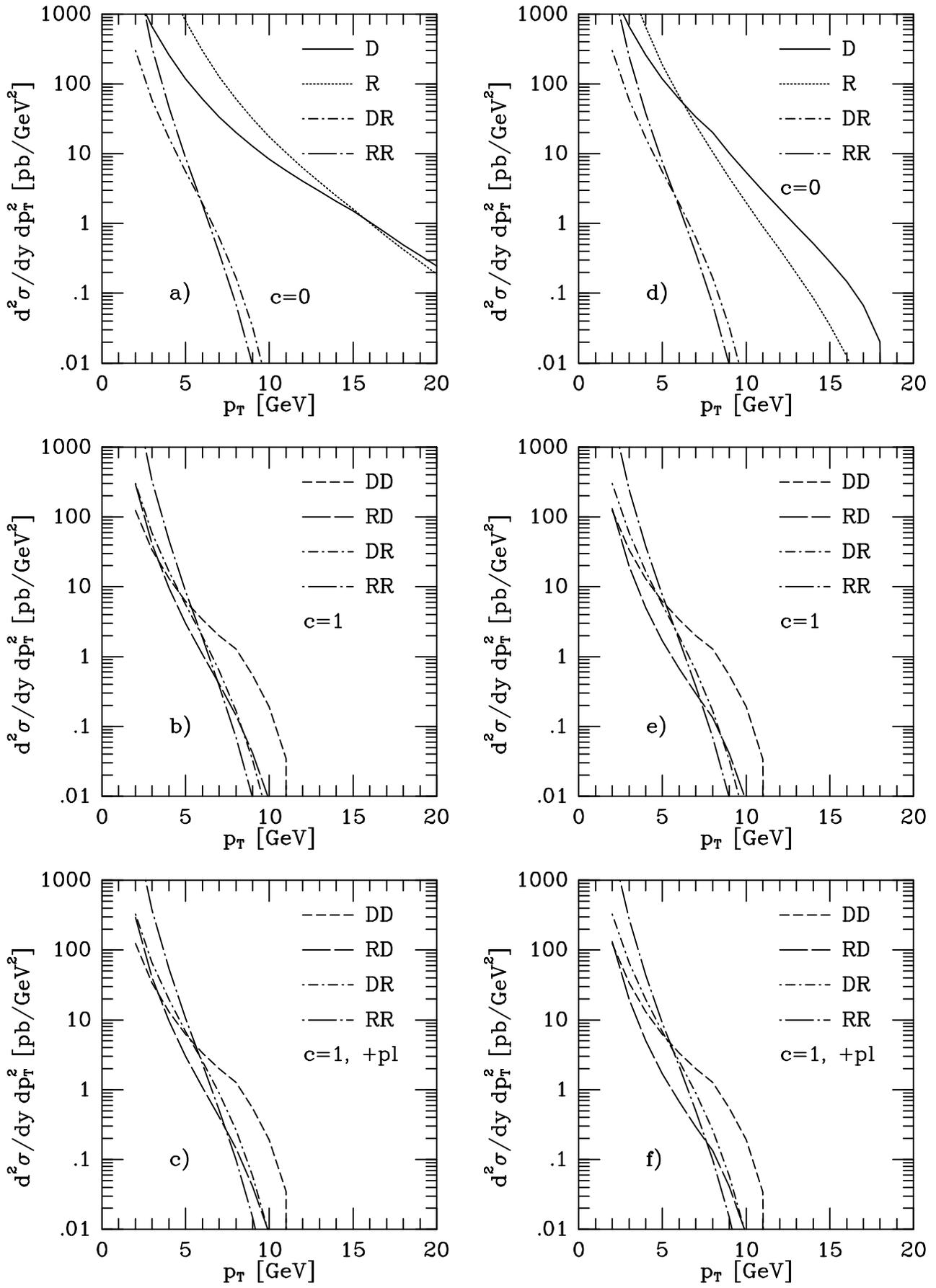
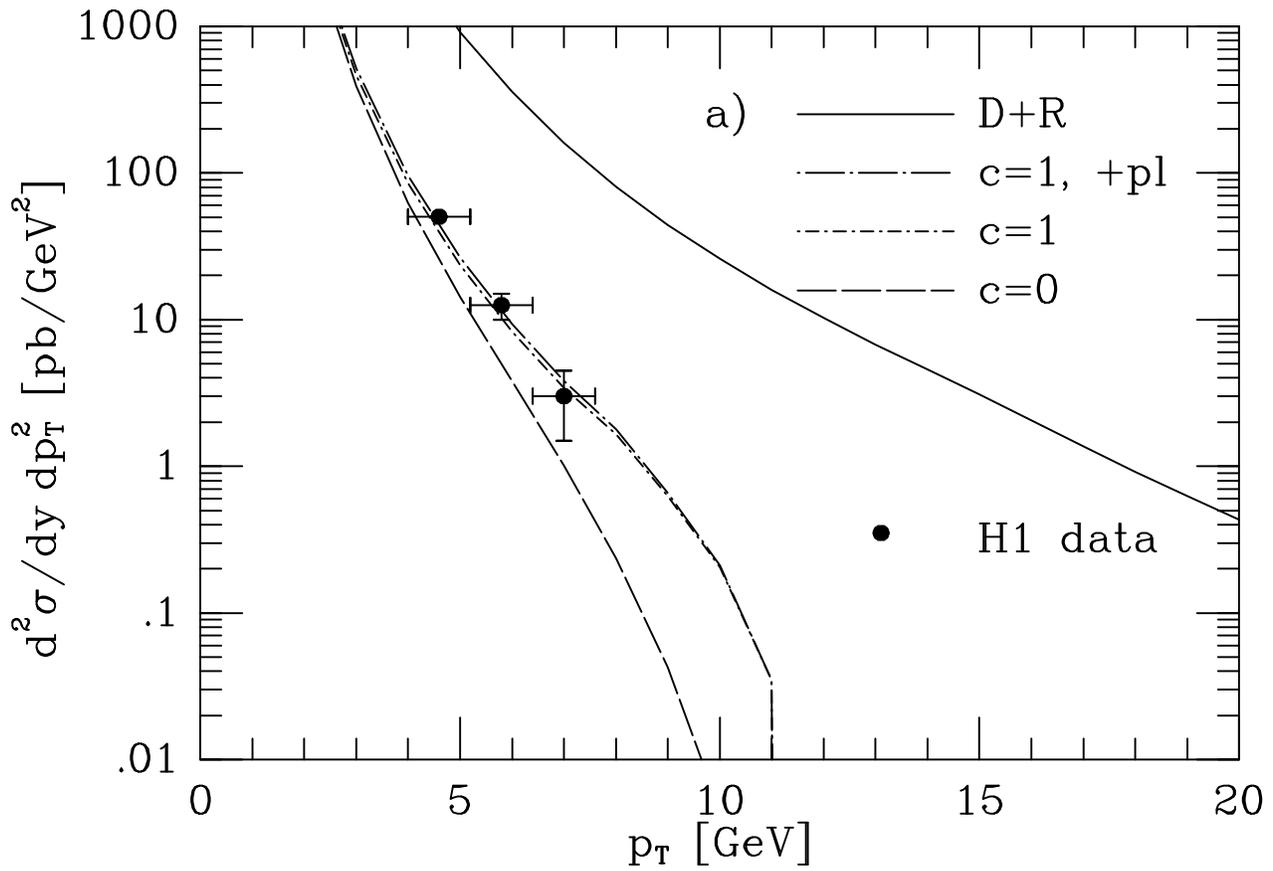
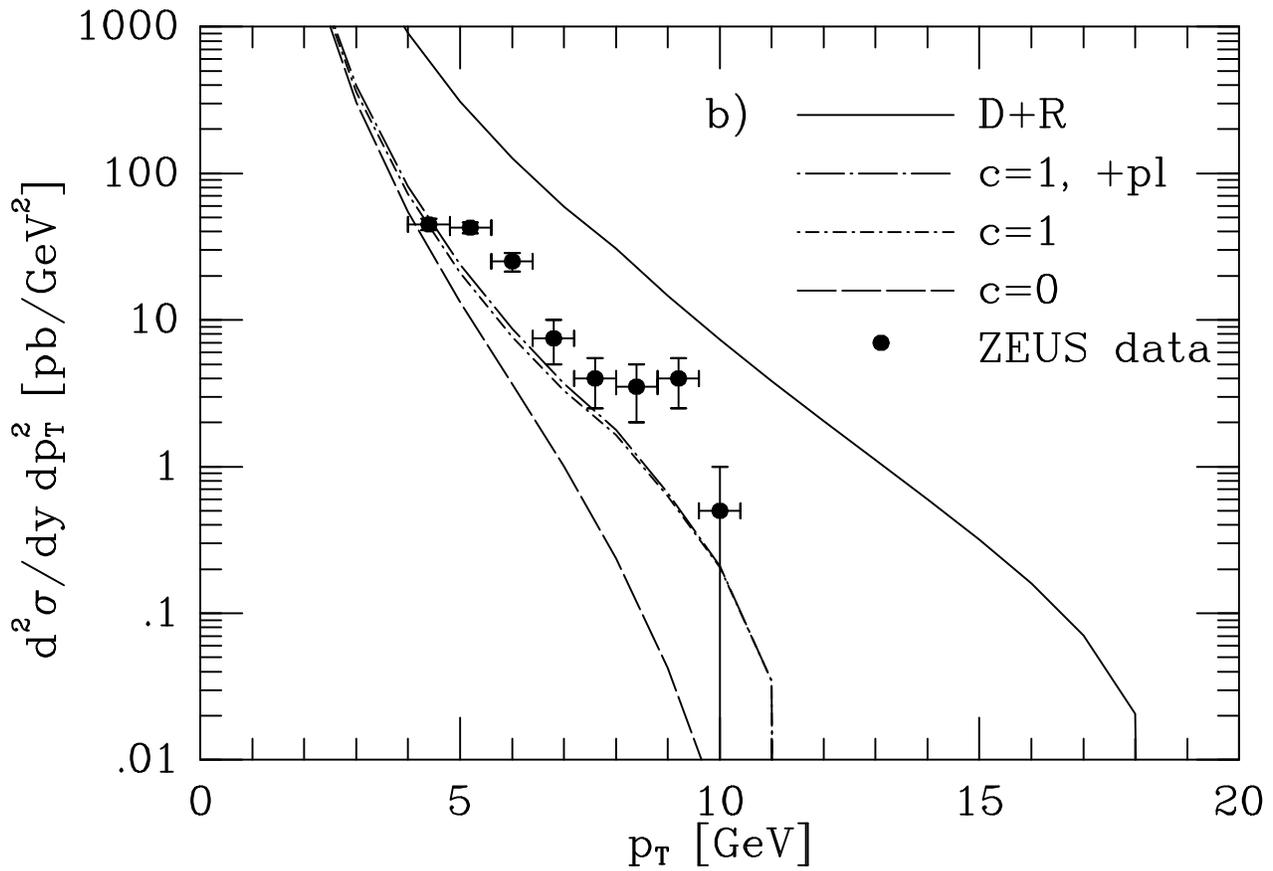
\end{document}